\documentclass[12pt]{article}
\usepackage{epsfig}

\def\beq{\begin{equation}}
\def\eeq#1{\label{#1}\end{equation}}
\def\eeqn{\end{equation}}

\def\beqa{\begin{eqnarray}}
\def\eeqa#1{\label{#1}\end{eqnarray}}
\def\eeqan{\end{eqnarray}}

\let\bar=\overbar

\def\O{{\cal O}}

\def\Dslash{\not{\hbox{\kern-4pt $D$}}}
\def\dslash{\not{\hbox{\kern-2pt $\del$}}}

\def\msb{{\bar{\ssstyle M \kern -1pt S}}}

\def\Title#1{\begin{center} {\Large {\bf #1} } \end{center}}

\begin{document}

\Title{New Physics Effects in $B$ Decays}

\bigskip\bigskip


\begin{raggedright}  

{\it Yuan CHAO\index{CHAO, Y.}\\
Department of Physics\\
National Taiwan University\\
10617 Taipei, TAIWAN}
\bigskip\bigskip
\end{raggedright}

\section{Introduction}

The two $B$-factories, Belle~\cite{Belle} and BaBar~\cite{BaBar},
has been played major roles in the
$B$ decays study. Their wonderful design and excellent operation enables
their fruitful analysis results. Also recently the Tevatron experiments,
CDF and D\O, join the game with their $B_S$ studies.

From the experimental results of $B$ decay studies, we learn that most of
measurements are consistent with the Standard Model (SM). One needs more
precise measurements, which relies on large statistics and good analysis tools,
to verify the theoretical predictions. Meanwhile, many unanticipated new
particles, like $X$, $Y$ and $Z$'s, are discovered as discussed in J.
Brodzicka \index{Brodzicka, J.} talk. After all, we still have some small
room for the New Physics. Some discrepancies from the SM has been found in
the measurements of the phases and magnitudes of CKM unitary triangle
\cite{CKM_KM, CKM_C}. There are also various theoretical models that possibly
give the contributions. These will relay on further validation with new
experimental results.

\section{Hints from the experiments}

The hints of discrepancies between data and standard model are found in
the following topics:

\subsection{Direct CP Violation}

\begin{figure}[th]
\begin{center}
\epsfig{file=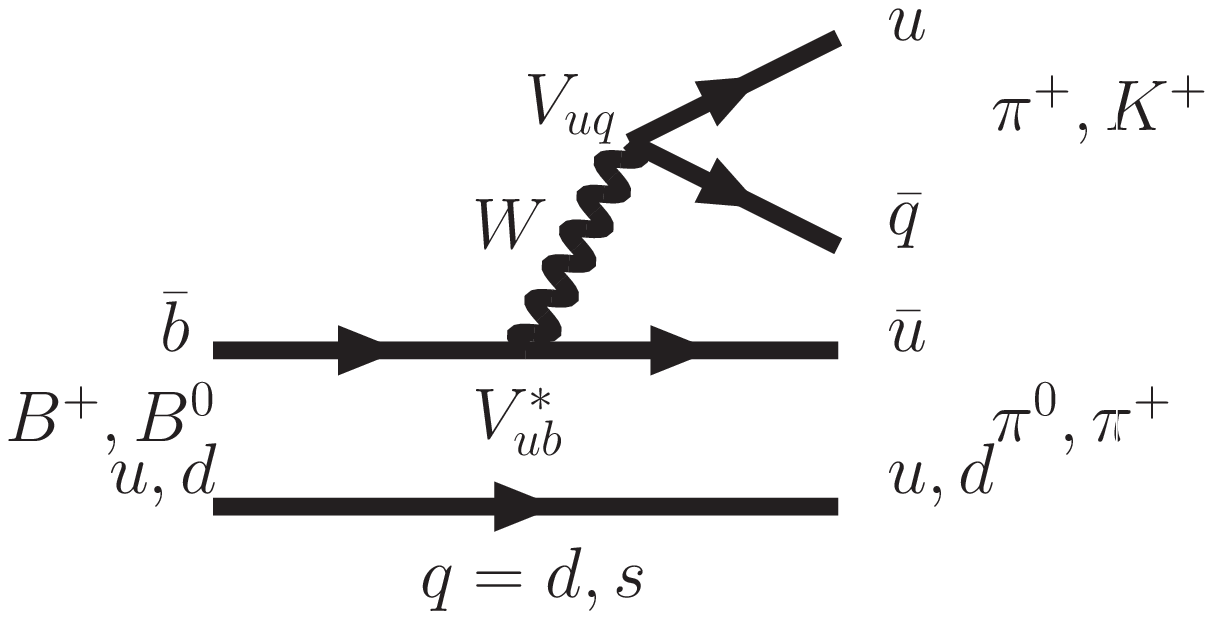,height=1.2in}
\epsfig{file=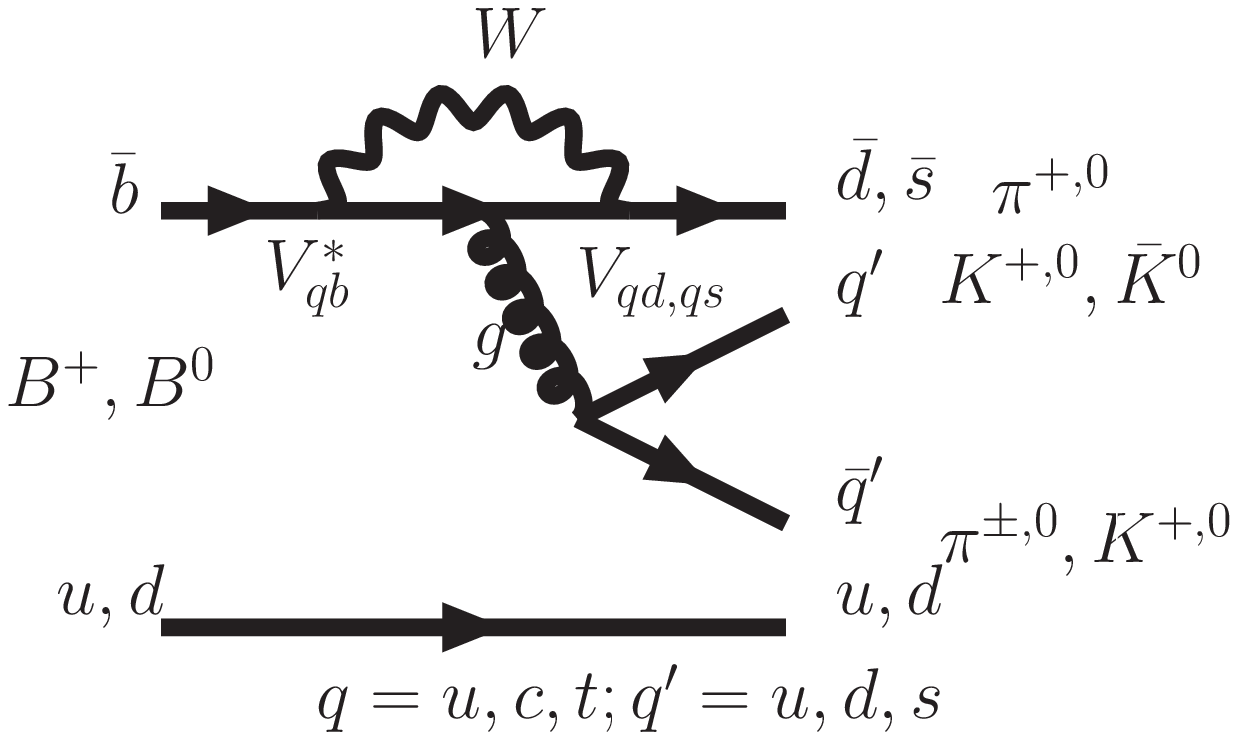,height=1.2in}
\caption{The Feynman diagrams of ``Tree'' and ``Penguin'' processes.}
\label{fig:tree_peng}
\end{center}
\end{figure}

In SM, $CP$ violation arises via the interference of at least two
processes with comparable amplitudes and difference $CP$ phases \cite{CKM_KM}.
The direct $CP$ violation (DCPV) of $B \to K\pi$ comes from the interference
of ``Tree'' and ``Penguin'' two major processes, shown in Fig.
\ref{fig:tree_peng}.

\begin{figure}[bh]
\begin{center}
\epsfig{file=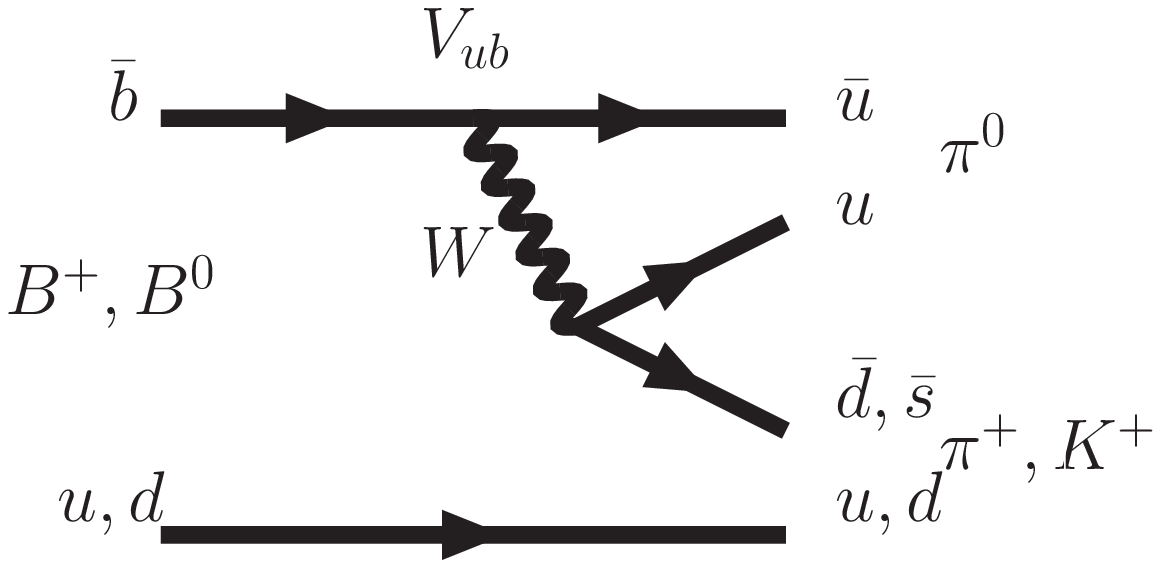,height=1.2in}
\epsfig{file=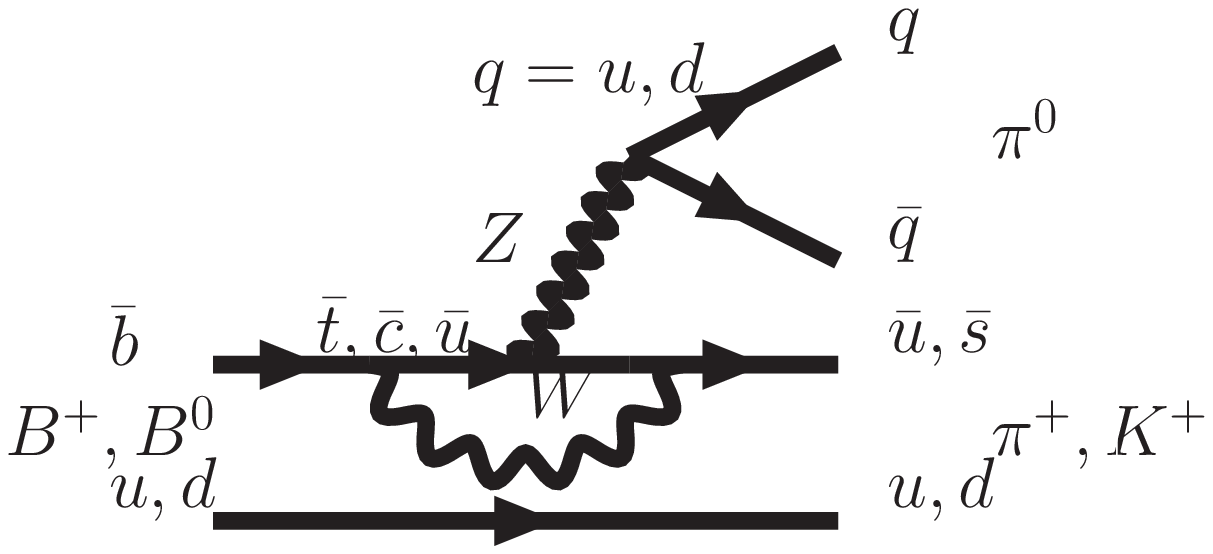,height=1.2in}
\caption{The Feynman diagrams of ``Color-suppressed Tree'' and
``Electroweak Penguin'' processes.}
\label{fig:ct_ewp}
\end{center}
\end{figure}

One would expect a similar DCPV of $B^\pm \to K^\pm\pi^0$ to $B^0 \to K^+\pi^-$.
However, from the $B$-factories experimental results, the difference is of
5.2 $\sigma$ significance with the world average measurements
\cite{BCP_B0, BCP_Bp}. This is once called the $K\pi$ puzzle \cite{Kpi}
as the other contribution processes of
$B \to K^+\pi^0$, ``Color-suppressed Tree'' and ``Electroweak Penguin'',
are theoretically expected to be small \cite{BCP_TH1}. Several theories
suggest ways to enhance these two contributions \cite{BCP_TH2}. However,
one needs experimental validations
on their predictions. A recent publication on {\it Nature} has a summary on
this issue and this is a non-concluded problem.

\subsection{Radiative and Electroweak Penguins}

The radiative $b \to s \gamma$ decays would be the most powerful modes
to constrain new physics. The deviation would be seen from their decay
rates. Experimentally, there are two methods to perform measurements:
fully inclusive and semi-inclusive, which sums up the exclusive channels.
Belle recently has an update with fully inclusive method and is the
current most precise measurement \cite{b2sg_belle}. From the comparison
of experimental world averages with next-next-leading-order (NNLO)
calculation \cite{b2sg_NNLO}, we find that the agreements of them has
been degraded. The most consist Andersen Gardi calculation has quite
large uncertainty. On the other hand, BaBar has a recent update with
the semi-inclusive method which sums up 16 fully reconstructed exclusive
modes \cite{b2sg_babar}. They also provide
a DCPV measurement with $-0.012\pm0.030(stat)\pm0.019(syst)$.

\subsection{Time-dept. CPV in $b \to s$ }

The time-dependent $CP$ violation (TCPV) measures the interference between
$B$ decays into final $CP$ eigen state and $B$ mixing into $\overline B$ and
decays into the same state. The indirect $CP$ violation has been established
in the $b \to c\bar c s$ with $B\to J/\psi K$ channel \cite{b2ccs_tcpv}.
However, the recent results of various $b \to s q\bar q$ channels shows
deviations with a na\"ive average \cite{b2sqq_tcpv}. Theoretically,
$b \to s q\bar q$ is through ``penguin'' process and has similar CP
values to $b \to c\bar c s$ process which is of ``tree'' process. As
there is no KM phase in $V_{ts}$, one would expect the same mixing induced
CP measurement. This deviation would imply some non-SM particles in the
loop of penguin process. The possible candidates would be the SUSY
particles or the K.K. particles \cite{KK} of extra dimension.
The current deviation of world average is about $2.2 \sigma$.

\subsection{Decays with Large Missing Energy}

The leptonic $B$ decays have sensitivity to new physics from charged
Higgs as long as the $B$ decay constant, $f_B$ is known. For example
the decay of $B \to \tau \nu$ can be expressed like this:
\[
   {\cal B}(B^+ \to \tau^+ \nu_\tau)= \frac{G^2_F m_B}{8\pi}m^2_\tau
(1-\frac{m^2_\tau}{m^2_B})^2 f^2_B |V_{ub}|^2\tau_B
\]

From the experimental point of view, the most sensitivity is from $\tau$
modes with 1-prong. The study is rather difficult as the interesting $B$
decays a single charged track and neutrinos which can't be seen. One needs
to utilize the information from the other $B$ pair produced in the
same event. The current measurements from Belle and BaBar are a combination
of $f_B \cdot |V_{ub}|$ which is of around $1.5\%$ level of uncertainty
in average \cite{b2taunu}. The difference between the measurements and
HPQCD calculation is within $1\sigma$ \cite{HPQCD}.

\subsection{New results from Tevatrons}

\begin{figure}[th]
\begin{center}
\epsfig{file=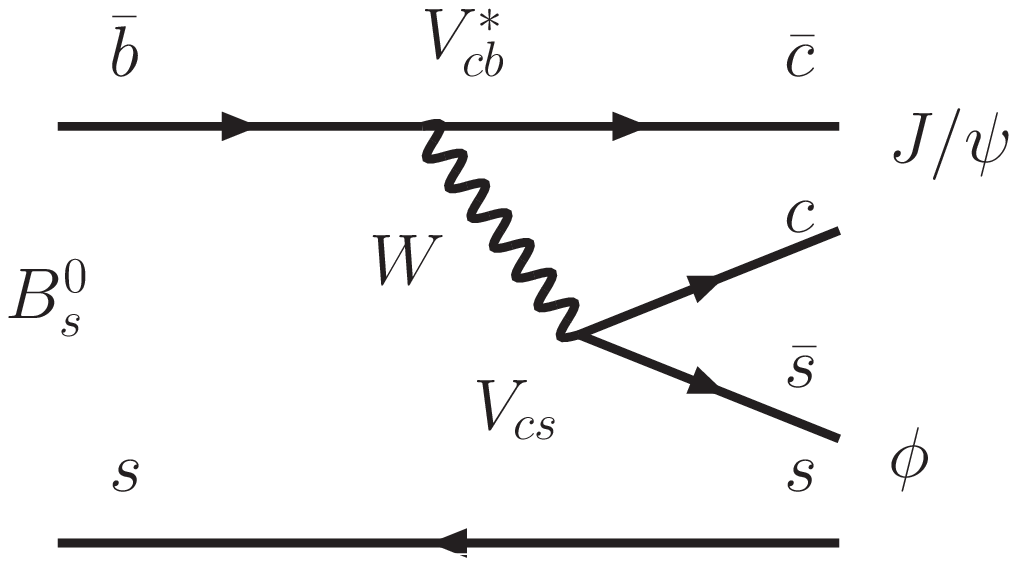,height=1.2in}
\epsfig{file=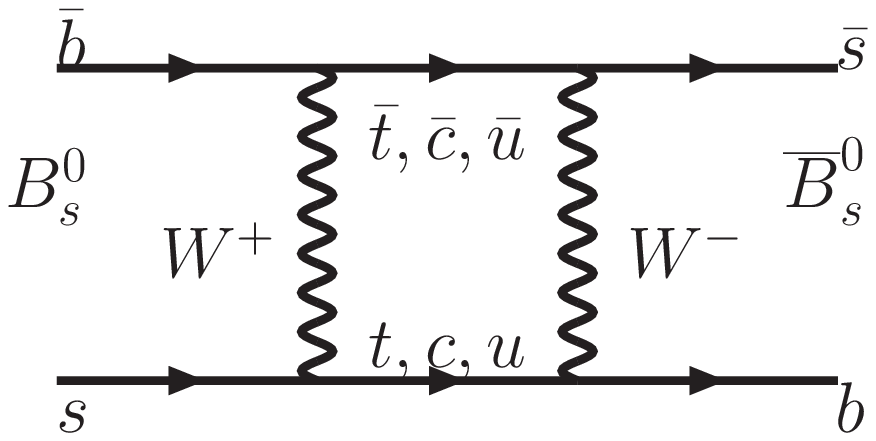,height=1.2in}
\caption{The Feynman diagrams of $B_S$ mixing processes.}
\label{fig:bs}
\end{center}
\end{figure}

The process of $B_S \to J/\psi \phi$ studied in the Tevatron experiments is
very similar to $B \to J/\psi K$ as shown in Fig.~\ref{fig:bs}.
However, the CP phase $\phi_{1~S}^{SM}$ (or $\beta_S^{SM}$) is
expected to be very small: $\phi_{1S}^{SM}=arg(-V{ts}V_{tb}^*/V_{cs}V_{cb}^*)
\sim 0.02$. Therefore, a non-zero measurement would be a hint of the effect
of new physics. Since $B_S$ is of spin 0 while $J/\phi$ and $\phi$ are of
spin 1. this leads to three angular momentum states that corresponds to
CP even and CP odd states. The large CP violation mixing coefficient
seen by CDF and D\O~indicates hints to new physics \cite{BS}.
Detailed explanation can be found in S. GIAGU's talk. \index{GIAGU, S.}

\section{Summary \& Conclusion}

The success of $B$-factories have brought us many fruitful physical results.
We also see some unexpected challenges to the SM. There are various
hints to new physics that have been pointed out in the previous paragraphs.
It's of no doubt that we still need more statistics to further clarifications.
Although the operation of BaBar has come to it's end early this Aprial,
people are now proposing upgrades to the present Belle while constructing a
new super $B$-factory. Of course, we are also looking forward to the up-coming
results in the LHC era.

\def\Discussion{
\setlength{\parskip}{0.3cm}\setlength{\parindent}{0.0cm}
     \bigskip\bigskip      {\Large {\bf Discussion}} \bigskip}
\def\speaker#1{{\bf #1:}\ }
\def\endDiscussion{}



 
\end{document}